\documentclass[prl,floatfix,twocolumn]{revtex4}
\usepackage{epsf}

\newcommand{\be}{\begin{equation}}
\newcommand{\ee}{\end{equation}}

\begin{document}

\title{The 
fourth root of the Kogut-Susskind determinant via infinite component fields.}

\author{ H. Neuberger}
\affiliation{
Rutgers University, Department of Physics
and Astronomy,
Piscataway, NJ 08855\\{\tt neuberg@physics.rutgers.edu}
}

\begin{abstract}
An example of interpolation by means of local field theories
between the case of normal Kogut-Susskind fermions and the case
of keeping just the fourth root of the Kogut-Susskind determinant
is given. For the fourth root trick to be a valid approximation
certain limits need to be smooth. The question about the validity
of the fourth root trick is not resolved, only cast into a local field
theoretical framework.
\end{abstract}

\maketitle

{\bf 1. Introduction. }

Recent simulations of QCD~\cite{milc} have been claimed to correctly include
sea quark effects by eliminating the extra tastes coming with 
Kogut-Susskind lattice fermions with the help of the so called ``fourth
root trick''. This trick amounts to replacing the local lattice
field theory, which would include all tastes, with one in which
the determinant of the gauge dependent Kogut-Susskind fermion matrix,
$K_s$, is taken at the power of $\frac{1}{4}$. 
The objective of this letter is to propose a class of embeddings
of the 4D lattice fermions into six dimensions, four of which are the
original lattice axes. These embeddings can be deformed by a parameter
of mass dimension, $\Lambda$, so that they look local from the four
dimensional viewpoint, so long as $\Lambda$ is of the order of the
inverse four dimensional lattice spacing $a$. Formally, if one takes
$\Lambda\to 0$ at fixed $a$ one recovers the fourth root trick. 
If one takes $\Lambda\to\infty$ at fixed $a$ one recovers 
a local theory with unmolested four tastes per species. 
Whatever scenario one has in mind for the validity of the fourth root
trick, it seems plausible that it should boil down to some robustness 
statement concerning the combined limits $a\to 0$ and $\Lambda \to 0$.

{\bf 2. General structure. }

The basic idea is a generalization of the work of Slavnov 
and Frolov~\cite{slav}. As is well known one could have proceeded from this
work alone to construct the overlap Dirac operator~\cite{overlap}, and
below I try to follow some of the steps that would have achieved this. 
However, the problem we are looking at here  
is substantially different, and by no means is it obvious what
the final conclusion (if any) about the validity 
of the fourth root trick would end up being. This letter is limited in scope
to merely setting the problem up in the 
language of infinite component fermi fields. 

We replace each original Kogut-Susskind fermion pair $\bar\chi, \chi$
by an infinite tower $\bar\chi^\alpha_n, \chi^\alpha_n$, 
labeled by indices $n,\alpha$, with the range of $\alpha$ being given by
a ``degeneracy'' $g_n$, for any given $n$. 
$n$ runs over all positive integers. We
are looking for a set of integers $g_n$, for which, formally at least, the
following holds: 
\begin{equation}
{\det}^{\frac{1}{4}} (K_s)=\prod_{n} {\det}^{g_n} (K_s)
\end{equation}
Neglecting questions of absolute convergence, we have the requirement
\begin{equation}
\label{a}
\sum_n g_n=\frac{1}{4}
\end{equation}
Obviously, with the $g_n$ all positive integers we can't have even conditional
convergence with the desired result. However, if we allow the statistics of
the fields to vary 
among the members of the tower, alternating signs might make~(\ref{a}) 
hold under conditional convergence. This is easily achieved by
\begin{equation}
\frac{1}{(1+x)^2}=\sum_{n=1}^\infty n(-1)^{n-1} x^{n-1},
\end{equation}
and setting $x=1$.

We can view the members of the towers as components of a vector in
an infinite Hilbert space. Each component is a vector in itself,
representing an ordinary Kogut-Susskind fermion. The infinite Hilbert
space is defined as the Hilbert space associated with a two dimensional
harmonic oscillator, whose Hamiltonian is $H$.
\begin{eqnarray}
&H=\frac{1}{2}{\vec p}^2+\frac{1}{2}{\vec q}^2=-\frac{1}{2}
\left ( \frac{\partial}{\partial{\vec q}}\right )^2+\frac{1}{2}{\vec q}^2 = 
a^\dagger a + b^\dagger b +1\cr
&[a^\dagger , a]=1=[b^\dagger ,b],~~[a,b]=[a,b^\dagger]=0\cr
&PaP=-a,~~PbP=-b,~~[H,P]=0\cr
&H|n;\alpha >=n|n;\alpha >,~n\ge 1,~~~g_n=n\cr
&a|1\rangle=b|1\rangle=0,~~P|n;\alpha\rangle = (-)^{n-1} |n;\alpha\rangle \cr
\end{eqnarray}
$P$ is the parity operation. We now declare any component consisting of an  
ordinary Kogut-Susskind structure to obey fermi statistics if it has
positive parity and bose statistics if its parity is negative. 
A component is fermionic if $n$ is odd and bosonic if $n$ is even.
Thus, the bosonic components can be fully paired up, providing a way
to make the path integrals over them convergent in spite of $K_s$ having
a spectrum that includes positive and negative values. 
The ground state of $H$ has eigenvalue 1, is non-degenerate and
labels a field of fermionic character. At a fixed $n>1$, the 
statistics is the same for all $g(n)$ vector components  
labeled by $\alpha$. If one deforms $H$,  
the deformation should preserve parity so that fields 
corresponding to different statistics do not get mixed. 
We denote the
inner product in the internal Hilbert space by $(,)$. With the fermion
action,
\begin{equation}
(\bar\chi, K_s \chi)=\sum_{n=1}^\infty \sum_{\alpha=1}^{g_n}
\bar\chi_n^\alpha K_s \chi_n^\alpha
\end{equation}
we formally have a local action where the entire contribution of
fermion loops is given by the fourth root of the determinant of $K_s$.

{\bf 3. Regularization. } 

We need to control the infinite number of fields; so far the
locality is a mere illusion, as we have an infinite number of massless
fields from the four dimensional viewpoint. This can be done by
giving all the higher members of the towers a large mass, of the order of the
ultraviolet cutoff affecting the four dimensional part of the fermion 
momenta, $\frac{1}{a}$:
\begin{equation}
K_s \rightarrow K_s +\Lambda f(H)
\end{equation}
where $f(x)> c >0$ for any $x=2,3,4....\infty$.
One can expand the sea quark contribution in Feynman diagrams which
would now contain also a trace over 
states in $H$. The convergence of that trace
would depend on the number of attached gauge field legs and the
asymptotic behavior of $f(x)$ as $x\to \infty$. It is clear that
demanding that $f(x)$ behave asymptotically as $x^\kappa$ will make
all diagrams converge if $\kappa$ is a large enough integer; for example
$\kappa=4$ is already an overkill, and $\kappa=2$ seems sufficient. 

The extra two ${\vec q}$ 
dimensions are seen only by the fermions; other fields are oblivious
of them. One could try to make up an operator $H$ which creates
a point-like defect at the origin of ${\vec q}$ space, so that
low energy-momentum fermionic modes are restricted to it.
In that case it would suffice to pick $f(x)\to c >0$ 
as $x\to\infty$. However, this is
not guaranteed to eliminate all ambiguities and some additional 
interpretation might be needed.

For a finite $\Lambda$, the limit $a\to 0$ will produce, most likely,
a theory with undesirable four-fold degeneracy for each 
fermion species if $f(1)=0$. 
Formally, if we take $\Lambda\to 0$ at fixed $a$, we get a model that
employs the fourth root trick. These observations reduce the problem to
an investigation of the combined limits $a\to 0$ and $\Lambda \to 0$.

{\bf 4. Discussion. }

It is sometimes argued that the fourth root trick is valid because (1), 
it agrees with experiment, and (2), it agrees with low energy predictions about
systems with approximate Goldstone bosons. Point (1) is well taken, and
the improvement of the agreement between lattice data and experimental 
data, in particular in cases where numerical data obtained from quenched
simulations showed distinct differences from experiment, is notable.
However, eventually, we would like numerical QCD to be so reliable that
when one detects a numerically significant 
discrepancy between its prediction and experimental 
data, one can interpret it as evidence of new physics.
In other words, we shouldn't  rely on experimental data when assessing
our calculations. Point (2) is not valid, as far as the author can see:
One could have a regime where low energy effective Lagrangians describe the
theory well, even if the ultraviolet completion of the theory is non-local.
At a more practical level, the number of free parameters in the effective
Lagrangian -- even ignoring (unjustifiably) some lattice Lorentz-violating
terms -- is too large to make the agreement credible to such a degree that
support for far reaching features can be abstracted from it. 

The fourth root trick has been recently criticized in~\cite{bunk} where
an operator corresponding to 
$K_s^{\frac{1}{4}}$ was considered. While 
this is one possible way to
get to a theory where the fermion contribution to the partition function is  
given by the fourth root of the determinant of $K_s$, it does not 
create an option for the system to become local, and therefore the
lack of locality one finds not necessarily implies that something is wrong
with the fourth root trick, as used. To substantially increase the 
amount of intuitive doubt about the fourth
root trick one would need to establish non-locality in a 
scheme that induces  
the system to select its true low energy modes in a smooth manner. 

Thus, it seems that the more theoretical and direct approach outlined above 
could be of use to those who feel strongly that the fourth root trick does
not mislead us. Alternatively, although {\it a priori} it 
would seem to be substantially more 
difficult to establish a negative, the present approach 
could yield a proof that the fourth
root trick never allows locality to be restored in the continuum limit. 
Obviously, the preferred outcome would be a positive one, and one would 
hope that
in the future the proponents of the fourth root trick would produce a proof 
of its validity 
using the above approach, or possibly some variation of it. 
The mere absence of
a real proof that the fourth root trick is inadequate 
will never provide satisfactory support for relying on numerical QCD 
carried out employing the fourth root trick.

{\bf 5. Acknowledgments.}

This research was partially supported  
by the DOE under grant number 
DE-FG02-01ER41165 at Rutgers University.


\end{document}